\documentclass[aps, prx, reprint, superscriptaddress]{revtex4-2}
\usepackage[utf8]{inputenc}
\usepackage{amsmath}
\usepackage{amsfonts}
\usepackage{amssymb}
\usepackage{hyperref}
\usepackage{graphicx}
\usepackage{epstopdf}
\usepackage{dcolumn}
\usepackage{bm}
\usepackage{xspace}
\usepackage{verbatim}
\usepackage[caption=false]{subfig}
\usepackage{chemformula}
\usepackage{booktabs}
\usepackage{chemformula}
\usepackage{orcidlink}
\usepackage{multirow}
\usepackage{color}

\hypersetup{colorlinks=true,linkcolor=blue,citecolor=blue, filecolor=blue,urlcolor=blue,breaklinks=true}

\begin{document}

\preprint{APS/123-QED}

\title{Fermionic Simulators for Enhanced Scalability of Variational Quantum Simulation} 

\author{Qingyu Li,\orcidlink{0000-0003-0269-8388}}
\affiliation{Institute of Fundamental and Frontier Sciences, University of Electronic Sciences and Technology of China, Chengdu 610054, China.}
\author{Chiranjib Mukhopadhyay,\orcidlink{0000-0002-4486-9061}}
\affiliation{Institute of Fundamental and Frontier Sciences, University of Electronic Sciences and Technology of China, Chengdu 610054, China.}
\author{Abolfazl Bayat,\orcidlink{0000-0003-3852-4558}}
\affiliation{Institute of Fundamental and Frontier Sciences, University of Electronic Sciences and Technology of China, Chengdu 610054, China.}

\begin{abstract}
Near-term quantum simulators are mostly based on qubit-based architectures. However, their imperfect nature significantly limits their practical application. The situation is even worse for simulating fermionic systems, which underlie most of material science and chemistry, as one has to adopt fermion-to-qubit encodings which create significant additional resource overhead and trainability issues. Thanks to recent advances in trapping and manipulation of neutral atoms in optical tweezers, digital fermionic quantum simulators are becoming viable. A key question is whether these emerging fermionic simulators can outperform qubit-based simulators for characterizing strongly correlated electronic systems.  Here, we perform a comprehensive comparison of resource efficiency between qubit and fermionic simulators for variational ground-state emulation of fermionic systems in both condensed matter systems and quantum chemistry problems. We show that the fermionic simulators indeed outperform their qubit counterparts with respect to resources for quantum evolution (circuit depth), as well as classical optimization (number of required parameters and iterations). In addition, they show less sensitivity to the random initialization of the circuit. The relative advantage of fermionic simulators becomes even more pronounced as interaction becomes stronger,  or tunneling is allowed in more than one dimension, as well as for spinful fermions. Importantly, this improvement is scalable, i.e., the performance gap between fermionic and qubit simulators only grows for bigger system sizes. 
\end{abstract} 

\maketitle


\section{Introduction}

The exponential growth of complexity by increasing the system size in quantum systems fundamentally limits the ability of classical computers in simulating quantum systems~\citep{feynman1982simulating,lloyd1996universal}. 
This makes quantum simulators indispensable to ensure continued technological and commercial development in material science \cite{ma2020quantum,PhysRevX.8.011044,PhysRevLett.120.110501}, chemistry \cite{hempel_quantum_2018,cao2019quantum,arguello2019analogue,bauer2020quantum}, and synthetic drug discovery \cite{Cavasotto2020}, among many others.
In all these fields, simulating the behavior of electrons, as fundamental elementary particles following fermionic statistics, and understanding their role in the formation of complex molecules is crucial. 
This makes the simulation of strongly-correlated fermionic systems of utmost importance~\cite{georgescu2014quantum,stanisic2022observing}. 
Quantum simulators are rapidly emerging in various platforms, including cold atoms in optical lattices~\cite{bloch2012quantum,gross_quantum_2017,yang2020observation, zhou2022thermalization}, ion-traps~\cite{ barreiro2011open,lanyon2011universal,blatt2012quantum,monroe2021programmable,kokail2019self}, superconducting devices~\cite{houck2012chip,arute2019quantum,google2020hartree,gong2021quantum,wu2021strong,mildenberger2022probing}, optical systems~\cite{aspuru2012photonic,wang2019boson,wang2020integrated,bao2023very,zhong2020quantum}, quantum dot arrays~\cite{byrnes2008quantum,hensgens2017quantum,mortemousque2021coherent} and Rydberg atoms~\cite{weimer2010rydberg,labuhn2016tunable,schauss2018quantum,Nguyen2018towards,keesling2019quantum,Omran2019Generation,browaeys2020many,scholl2021quantum,cohen2021quantum,young2022tweezer,yan2022two-dimensional,wu2021concise,cheng2023variational}. 
Most of these quantum simulators are qubit-based as, in principle, they can eventually achieve universal quantum computation in a digital manner~\cite{lloyd1996universal}. 
However, in the absence of error correction, Noisy Intermediate Scale Quantum (NISQ) simulators are far from being perfect and thus achieving near-term universal quantum computation is not foreseeable~\cite{Preskill2018quantumcomputingin, bharti2022noisy}. 
Quantum supremacy of NISQ simulators over classical computers has been demonstrated for specific problems \cite{arute2019quantum,zhong2020quantum,wu2021strong,gong2021quantum}. 
However, these problems have no direct practical applications.
Thus, a key open problem is whether NISQ simulators can achieve a practical quantum advantage over their classical counterparts~\cite{daley2022practical}. 

In NISQ era, hybrid quantum-classical variational algorithms are seen as the most promising route to demonstrate supremacy over fully classical computation paradigm \cite{peruzzo2014variational,cerezo_variational_2021}. 
In these algorithms, the target outcome is written variationally in terms of a minimum cost function which is measured as the output of a parameterized quantum circuit. 
The measured cost function is fed into a classical optimizer to be minimized which updates the parameters of the circuit. By iterating the algorithm eventually, the cost function reaches its minimum. Indeed, dividing the complexity between a quantum circuit and a classical optimizer allows a fairly shallow quantum circuit to potentially achieve a quantum advantage. Thus, a complete resource analysis of any variational quantum algorithm necessarily involves analysis of both quantum and classical resources. 
Quantum resources can be 
quantified through circuit depth or equivalently the number of two-qubit gates. Classical resources, however, quantify the complexity of the optimization problem through the number of required iterations as well as the number of optimizable parameters. A lot of efforts have been dedicated to enhancing the efficiency of variational quantum algorithms by saving both quantum and classical resources through inventing error mitigation methods~\cite{temme_error_2017,sagastizabal2019experimental,ravi2022vaqem,PhysRevX.8.031027,qin2022overview}, efficient design of quantum circuits~\cite{liu2019variational,huang2022robust}, exploiting symmetries~\cite{Kazuhiro2020Symmetry,gard2020efficient,Lyu2023symmetry,han2023multi} and accelerating the classical optimizer~\cite{lyu2020accelerated,alam2020accelerating,sweke2020stochastic,muller2022accelerating}. 

Variational Quantum Eigensolver (VQE)~\cite{peruzzo2014variational,wang2019accelerated, wu2021expressivity,tilly2022variational,uvarov2020variational} is perhaps the most widely investigated variational quantum algorithm which has been demonstrated experimentally for both condensed matter systems~\cite{ 
sagastizabal2021variational,kokail2019self,bravo2020scaling} and quantum chemistry problems~\cite{kandala2017hardware,li2019variational,google2020hartree,metcalf2020resource,cai2020quantum,zhang2021shallow, delgado2021variational,yoshikawa2022quantum}. The goal of the VQE algorithm is to prepare an individual eigenstate, e.g. the ground state, of a given Hamiltonian. For the most basic version of the VQE, the average energy is used as the cost function whose minimization results in the ground energy of the system. Thus, the output of the optimal circuit represents the ground state.  The conventional qubit-based quantum simulators cannot directly simulate fermionic systems. Certain mathematical transformations, such as Jordan-Wigner~\cite{jordan1993paulische,lieb1961two,batista2001generalized,derzhko2001jordan} or Bravyi-Kitaev~\cite{bravyi2002fermionic}, are needed to map  fermionic operators into Pauli strings~\cite{stanisic2022observing}. The resulting qubit Hamiltonian might be highly non-local, which makes the VQE algorithm more costly due to increased required measurement and makes it more susceptible to Barren plateau~\cite{mcclean2018barren,wang2021noise,uvarov2021barren,cerezo2021cost, kim2021universal, larocca2023theory} which significantly slows down the training. 
Recently, the network of trapped atoms in optical tweezers has been exploited to realize an analog Fermi Hubbard quantum simulator~\cite{norcia2018microscopic,yan2022two,spar2022realization}. In a recent proposal~\cite{gonzalez2023fermionic}, such systems have been proposed for realizing a fermionic \emph{digital} quantum simulator in which two-particle quantum gates can be performed by exciting the atoms to Rydberg states. This idea can also be utilized for simulating gauge field theories~\cite{gonzalez2022hardware,zache2023fermion}. Indeed, the feasibility of fermion-based quantum simulators opens huge possibilities for simulating fermionic systems without the overhead cost of the transformation from fermionic operators to Pauli strings.
A quantitative analysis is indeed essential to find out whether fermionic simulators can make VQE simulations more resource-efficient. If so, how this resource efficiency depends on the geometry of the fermionic system and scales with the system size.


The outline of the paper is as follows. We briefly review the background material for this work in the rest of this section, basics of the VQE algorithm are introduced in \ref{sec:VQE}, followed by their implementations for qubit (\ref{sec:VQE_Qubit_Simulator}) and fermion based (\ref{subsec:vqe_fermionic_simulator}) simulators. We detail our circuit design and numerical simulation techniques in Sec.~\ref{sec:II}. Sec.~\ref{sec:III} contains the results for variational simulation of the spinless Fermi-Hubbard model with fermionic simulators in 1D open chain (\ref{subsec:spinless_open_chain}), ladder (\ref{subsec:spinless_ladder}), and 2D lattice (\ref{subsec:spinless_2d}) configurations. This is extended to the spinful Fermi-Hubbard model in Sec.~\ref{sec:IV} with 1D open chains (\ref{subsec:spinful_open}), and ladder (\ref{subsec:spin_ladder}) configurations. \textcolor{black}{The scalability of classical and quantum resource requirements for ground-state finding with fermionic and qubit simulators is analyzed in Sec.~\ref{sec:scalability}}.  Results for Fermionic simulation for the water molecule are presented in Sec.~\ref{sec:V}, before we conclude in Sec.~\ref{sec:conclusion}. 

. 

\section{Variational Quantum Eigensolver Algorithm}
\label{sec:VQE}

In this section, we briefly review the Variational Quantum Eigensolver (VQE), as one of the most widely used near-term algorithms in the field of quantum simulation. VQE has been developed to determine individual eigenvalues of a many-body system and prepare their corresponding eigenstates.  The simplest version of the VQE is for simulating the ground state and is built on the Ritz variational principle, i.e., if one chooses a trial state $|\psi (\vec{\theta})\rangle {=} \mathcal{U}(\vec{\theta}) |\psi_0\rangle$ in which $\mathcal{U}(\vec{\theta})$ is a unitary operator parametrized by $l_p$ real parameters $\vec{\theta} {=} \lbrace \theta_1, \theta_2, ...., \theta_{l_p} \rbrace $, then the ground state energy is bounded by average energy, namely  
\begin{equation}
    E_0 \leq \langle \psi (\vec{\theta})|H|\psi (\vec{\theta})\rangle.
\end{equation}
This implies that if $|\psi (\vec{\theta})\rangle$ is expressible, i.e. includes the ground state for a certain choice of  $\vec{\theta}{=}\vec{\theta}_{opt}$, then by minimizing the average energy, as the cost function, one can get the ground state energy as $E_0{=}\langle \psi (\vec{\theta}_{opt})|H|\psi (\vec{\theta}_{opt})\rangle$. In this case, the output of the simulator corresponds to the ground state, namely $|E_0\rangle{=}|\psi (\vec{\theta}_{opt})\rangle$. The parameterized quantum state $|\psi (\vec{\theta})\rangle$ can be obtained through the operation of a parameterized quantum circuit $\mathcal{U}(\vec{\theta})$. Theoretically, a quantum analog of the universal approximation theorem has been developed, which indicates that sufficiently deep quantum circuits can approximate any target function to the desired accuracy \cite{wu2021expressivity}. However, the practically relevant problem of designing the shallowest possible quantum circuit for simulating the ground state of a Hamiltonian remains an area of intensive research. This includes incorporating the symmetries of the system into the design of the circuit~\cite{gard2020efficient, meyer2023exploiting,Lyu2023symmetry} or exploiting evolutionary~\cite{chivilikhin2020mog,huang2022robust} and machine learning~\cite{ostaszewski2021reinforcement,NEURIPS2021_97244127,PRXQuantum.2.010324} algorithms for simplifying the circuit. 

As mentioned before, VQE algorithm relies on two different types of resources: (i) quantum resources which can be quantified through the depth of the quantum circuit or equivalently the number of two-qubit gates $R_\text{Q}$; and (ii) classical resources which quantifies the complexity of the classical optimization through
\begin{equation}
    R_\text{C}=l_\text{p}\times  l_\text{I}
\end{equation}
where $l_\text{I}$ is the average number of iterations that the optimizer needs to iterate for converging to a given precision. The resource efficiency, namely minimizing both $R_{\text{Q}}$ and $R_\text{C}$, is essential for the scalability of the VQE algorithm and achieving a quantum advantage in NISQ era.

\subsection{VQE Algorithm for Fermionic Systems on Qubit-Based Quantum Simulators }
\label{sec:VQE_Qubit_Simulator}

In this paper, we focus on simulating the ground state of fermionic many-body systems in both condensed matter physics and quantum chemistry. However, most of the quantum simulators that are available are qubit-based which do not satisfy Fermi statistics. This means that to implement VQE on such systems, one has to first map the fermionic Hamiltonian into a qubit one. The spinless fermionic annihilation and creation operators acting on-site $j$, represented by $c_j$ and $c_j^\dagger$ respectively, follow the Fermi statistics as 
\begin{eqnarray}
    \{c_j,c_k\}&=&0 \cr
    \{c_j,c_k^\dagger\}&=&\delta_{jk}
\end{eqnarray}
where $\delta_{jk}$ is the Kronecker delta function, and $\lbrace ~,~ \rbrace $ is the anti-commutator. There are several methods to map these fermionic operators into qubit basis. The Jordan-Wigner transformation is most prominent among such maps \footnote{Although other qubit-fermion mappings have been proposed recently. See, e.g., Ref~\cite{Steudtner2018fermion} }, and relates Fermion operators $\lbrace c_{j}\rbrace$ to Pauli spin operators via the following mapping
\begin{eqnarray}
    c_{j}^{\dagger} & = &\left(\otimes_{k=1}^{j-1}\sigma_k^z\right) \sigma_{j}^{+} \\
    c_{j} & = &\left(\otimes_{k=1}^{j-1}\sigma_k^z \right) \sigma_{j}^{-} \\
    n_j&=&c_{j}^{\dagger}c_{j} = \frac{\mathbb{I}-\sigma_{j}^{z}}{2},
\end{eqnarray}
where $n_j$ is the number operator, $\sigma^\alpha_j$ (with $\alpha{=}x,y,z$) is the Pauli operator acting on site 
$j$ and $\sigma^\pm_j{=}(\sigma^x_j\pm i\sigma^y_j)/2$. Note that the string operator $\otimes_{k{=}1}^{j-1}\sigma_k^z$ is highly non-local which creates multi-body interaction in the qubit basis. In particular, when the fermionic Hamiltonian contains long-range tunnelings or describes high-dimensional lattices, the corresponding qubit Hamiltonian becomes highly non-local. The emergence of such non-local terms puts extra resource overheads on VQE simulation and creates several drawbacks, including (i) the number of measurements for estimating average energy increases; (ii) the optimization becomes more susceptible to the Barren plateau phenomenon due to non-local terms in the cost function which makes the convergence slower~\cite{mcclean2018barren,cerezo2021cost}; and (iii) the circuit design becomes challenging and quite arbitrary. Indeed, if one could bypass the Jordan-Wigner transformation step for mapping fermions to qubits, then the above inefficiencies would be prevented, resulting in a more efficient simulation of fermionic quantum systems.

\subsection{VQE simulation on Fermion-Based Quantum Simulators}
\label{subsec:vqe_fermionic_simulator}

In the qubit quantum computational paradigm, the qubits are assumed to be distinguishable. On the other hand, elementary particles or atoms, either Bosons or Fermions, are completely indistinguishable. Thus, for Fermionic or Bosonic computation models, the natural approach is to consider distinguishable localized energy modes on a similar footing as qubits. For Bosons, this leads to elementary gate operations being linear optical unitaries augmented with some non-Gaussian operation like a photon detector, being universal \cite{knill2001scheme}. For Fermions, the situation is more interesting, as projective measurement on Fermion modes does not achieve universality when coupled with free-Fermionic quadratic unitaries \citep{terhal2002classical}. Importantly for our purpose, Bravyi and Kitaev~\citep{bravyi2002fermionic} obtained the following set of Fermionic unitaries which are universal when restricted to global particle-number conserving transformations
\begin{eqnarray}
    \mathcal{BK} =   \lbrace e^{\frac{i \pi}{4} n_{j}}, e^{\frac{i \pi}{4} n_{j} n_{j'}},
    e^{\frac{i \pi}{4} \left( c_{j}^{\dagger} c_{j'} +  c_{j'}^{\dagger} c_{j} \right) }  \rbrace,
    \label{eq:bravyi-kitaev}
\end{eqnarray}
Physically, the first unitary acts on a single site and encodes information about particle numbers for each mode, the second unitary signifies interaction between fermions in different sites, and the last unitary signifies particle hopping between sites. This is an interesting feature that makes fermions very distinct from qubits. While in qubit systems, one type of two-qubit operation, e.g. controlled-not, is enough for universal computation, in fermionic quantum simulators one needs two different types of two-particle operations. 

Recently, a novel approach for developing digital fermionic quantum simulators in neutral atom arrays with interactions mediated by Rydberg states has been proposed~\cite{gonzalez2023fermionic}. In fact, the authors argue that the following generalization of the Bravyi-Kitaev gate set can be precisely realized in experiments. 
\begin{equation}
\mathcal{G} = \lbrace \mathcal{U}_{jj'}^{\text{tun}} (\vec{\theta}), \mathcal{U}_{jj'}^{\text{int}} (\theta) \rbrace
\end{equation}
in which
\begin{eqnarray}
    \mathcal{U}_{jj'}^{\text{tun}} (\vec{\theta}) &=& e^{-i \left[ \frac{\theta_1}{2}\left( e^{-i\theta_2} c_{j} c_{j'}^{\dagger}+ e^{i\theta_2} c_{j'} c_{j}^{\dagger} \right) + \frac{\theta_3}{2} (n_j - n_j') \right]}, \cr
    \mathcal{U}_{jj'}^{\text{int}} (\theta) &=& e^{-i\theta n_{j} n_{j'}}
    \label{eq:gen_bk}
\end{eqnarray}
where, $\mathcal{U}_{jj'}^{\text{tun}}$ is a tunneling gate allowing fermions to shift at different sites, and $\mathcal{U}_{jj'}^{\text{int}}$ is an interaction gate which
gives relative phases between different charge configurations.  
Thus, the fermion systems can be simulated using these two gates on fermionic quantum simulators directly, instead of being implemented indirectly on a qubit-based simulator by using the Jordan-Wigner transformation with the additional overhead of quantum resources.

The digital fermionic quantum simulator in Ref.~\cite{gonzalez2023fermionic} exploits fermionic atoms trapped in optical tweezers. 
In order to implement the interaction and tunneling gates $\mathcal{U}_{jj'}^{\text{int}}$ and $\mathcal{U}_{jj'}^{\text{tun}}$ one can excite the atoms into Rydberg states. In particular, tunneling gates can be realized by different methods, two of which are critical, the MERGE protocol and the SHUTTLE protocol.
In the MERGE method~\cite{kaufman2014two, spar2022realization,young2022tweezer}, two nearby optical traps are brought so close together that atoms can tunnel through to each other and then separate these two traps. 
In this case, the parameters $\vec{\theta}$ of $\mathcal{U}_{jj'}^{\text{tun}}$ are determined by the behavior of tweezers and the custom-designed merging and splitting protocols. 
On the other way, namely the SHUTTLE approach, two tweezer arrays are employed each trapping different spin states of the atom~\cite{daley2008quantum}. Then, when these tweezers overlap, Rydberg-mediated level transitions in the spin subspace also lead to coherent delocalization of positional degrees of freedom. This delocalized position wave function can then be collapsed in the new site of choice, making it possible to implement the $\mathcal{U}_{jj'}^{\text{tun}}$.
The interaction gate $\mathcal{U}^{\text{int}}_{jj'}$ can be similarly implemented by bringing the corresponding optical tweezers within Rydberg blockade distance and then driving a laser pulse to create a Rydberg excitation that couples to the internal angular momentum degrees of freedom. By suitably choosing the shape of the laser pulse,  $\mathcal{U}^{\text{int}}_{jj'}$ \cite{levine2019parallel} can be obtained.
For unrelated site pairs, both $\mathcal{U}^{\text{tun}}_{jj'}$ and $\mathcal{U}^{\text{int}}_{jj'}$ can be parallelized, i.e., implemented simultaneously all across the length of an array of neutral atoms.

\section{Circuit Design for Both Qubit and Fermion Quantum Simulators}
\label{sec:II}
\begin{figure}
    \centering
    \includegraphics[width=0.5\textwidth]{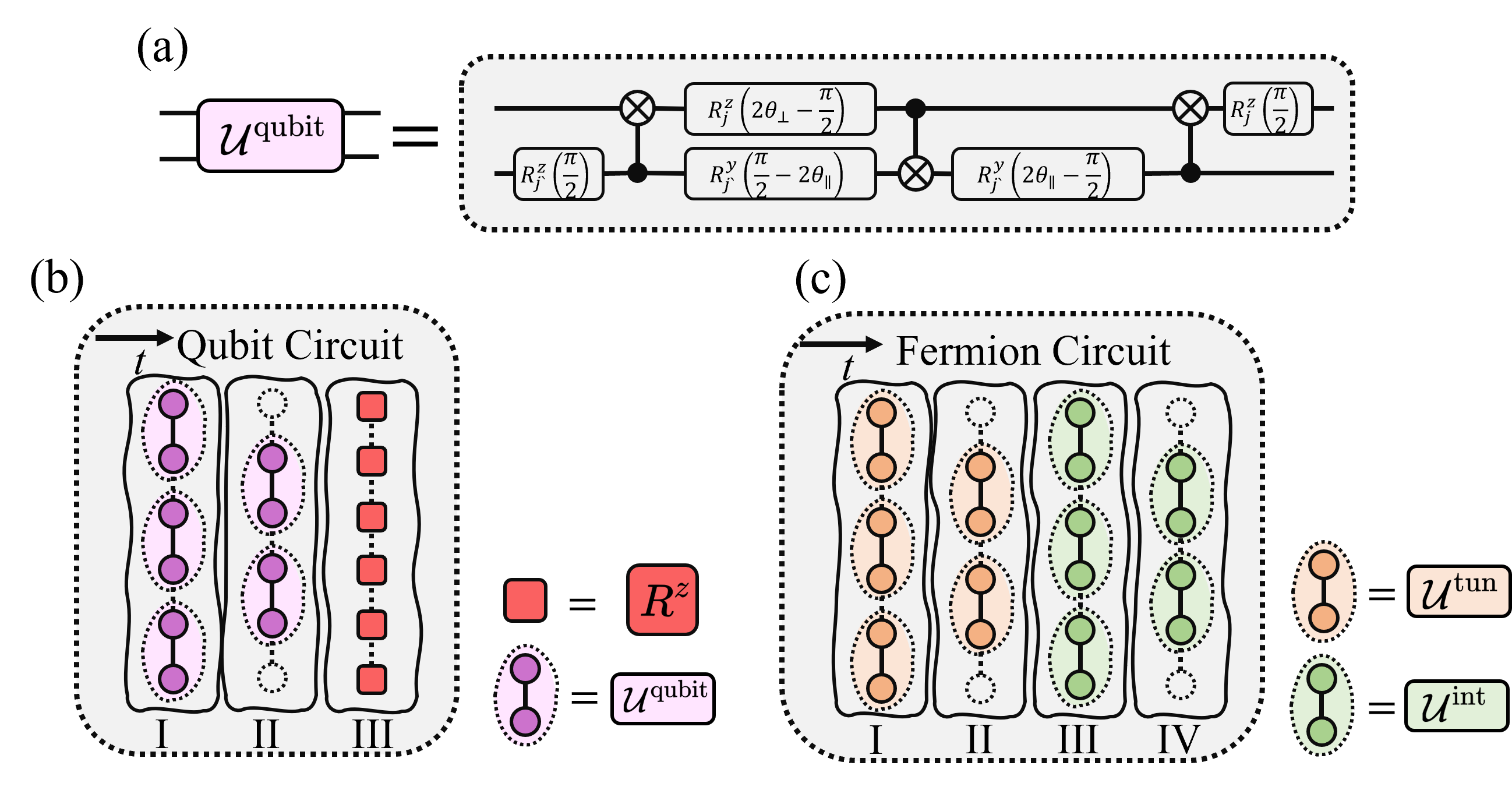}
    \caption{(a) Decomposition of two-qubit gates $\mathcal{U}_{jj'}^\text{qubit}$ (purple) in terms of rotations and CNOT gates, (b) Decomposition of one layer of qubit circuit used in this paper, first $\mathcal{U}_{jj'}^\text{qubit}$ (purple) is applied on all neighboring sites and then $R^z_j(\theta)$ (red) is applied on all sites, (c) Decomposition of one layer of fermionic circuit used in this paper, first $\mathcal{U}_{jj'}^{\text{tun}}$ (orange) is applied on all neighboring sites and then $ \mathcal{U}_{jj'}^{\text{int}}$ (green) is applied on the same way.}
    \label{fig:circuit_design}
\end{figure}

In this paper, we focus on Hamiltonians in which the number of fermions is conserved. By mapping to qubit basis, the resulted Hamiltonian $H_q$ preserves the number of excitation, namely $[H_q,S^z]{=}0$ where $S^z{=}1/2\sum_j \sigma_j^z$. To incorporate this symmetry in the circuit design we rely on two-qubit gates $\mathcal{U}_{jj'}^\text{qubit}$ and single qubit rotations $R^z_j(\theta)$ which take the form
\begin{eqnarray} 
\mathcal{U}_{jj'}^\text{qubit}&=&e^{+i\left[\theta_{\parallel}(X_jX_{j'}+Y_jY_{j'})+\theta_\perp Z_jZ_{j'}\right]},\cr
R^z_j(\theta)&=&e^{-i\frac{\theta}{2}\sigma_j^z}.
\label{eq:xyz}
\end{eqnarray}
In Fig.~\ref{fig:circuit_design}(a), we depict the quantum circuit which realizes the two-qubit operation $\mathcal{U}_{jj'}^\text{qubit}$. By combining these gates one can construct a quantum circuit that naturally preserves the number of particles. One layer of such a circuit is shown in Fig.~\ref{fig:circuit_design}(b). In order to enhance the expressivity, several layers of the circuit are concatenated to make a deeper circuit.  

For designing the circuit in fermionic digital quantum simulators we can use different arrangements of $\mathcal{U}^{\text{tun}}_{jj'}$ and $\mathcal{U}^\text{int}_{jj'}$. We have found out that a particular arrangement performs better for the VQE simulation. In this advantageous configuration, in a given circuit layer,  all the tunneling gates $\mathcal{U}^{\text{tun}}_{jj'}$ are grouped together to act on all the bonds and then followed by similar arrangements for interaction gates  $\mathcal{U}^\text{int}_{jj'}$. In Fig.~\ref{fig:circuit_design}(c), we depict one layer of a typical circuit for a one-dimensional system. 

Note that in the following section, the qubit and fermion circuits have a little difference as shown in Fig.~\ref{fig:circuit_design}, but keep the same gate arrangements. Initially, the circuit parameters of both VQEs are randomly set to near $0$ to ensure $\langle H \rangle$ of both VQEs start from similar values. 

For the classical optimization part, we use the Broyden-Fletcher-Goldfarb-Shanno (BFGS) algorithm \cite{broyden1970convergence,fletcher1970new,goldfarb1970family,shanno1970conditioning}, and restrict maximum iterations $l_\text{I}$ to $150$ to minimize the average energy $\langle H \rangle$, which is sufficient in most cases. It is worth mentioning that other classical optimizers also work and do not qualitatively affect the conclusions.  In addition, for each case, we repeat the procedure ~$100$ times for random initializations, over which the results are averaged to be statistically meaningful. 

\textcolor{black}{It is worth emphasizing that depending on the Hamiltonian to be simulated, the above qubit circuit design may not be the most efficient one. For example, unitary coupled cluster ansatz-based circuits are more favored for molecular simulations. In particular, an adaptive optimization of circuit design~\citep{Grimsley2019An}, has become very popular in recent years. 
We stress that this technique remains equally available for fermionic simulators too, with the \emph{operator pool} of Ref.~\cite{Grimsley2019An} being the fermionic gates in Eq.~\eqref{eq:gen_bk}. However, for the sake of fair comparison between qubit and fermionic quantum simulators, we use the same strategy, i.e. adopting symmetry-preserving circuits and using the same classical optimizer, for both qubits and fermions. Indeed, one can use more complicated strategies, such as adaptive optimization of the circuit design~\citep{Grimsley2019An}, in both cases and improve the performance. Here, our objective is to provide a comparison of the performance of the qubit and the fermionic simulators and thus these improvements are out of the scope of the paper and left for future investigations.}


\section{VQE Simulation of spinless Fermionic systems}
\label{sec:III}
Now we start the analysis of Fermionic quantum systems. In this section, we neglect the explicit contribution of spin degrees of freedom and show how VQE simulation of such systems on a fermionic digital quantum simulator can outperform 
the performance of qubit-based quantum simulators. We focus on the simplest VQE case, namely finding the ground energy of spinless fermionic Hamiltonians. In the absence of spin degrees of freedom, the Pauli exclusion principle dictates that there can only be one electron per site. Therefore, a typical spinless fermion Hamiltonian reads as
\begin{equation}
    H(t,V,\mu)=-t\sum_{\langle jj'\rangle}^{N}(c_j^\dagger c_{j'} +\text{h.c.})+V\sum_{\langle jj'\rangle}n_jn_{j'}-\mu\sum_jn_j,
\label{eq:hubbard_spinless}
\end{equation}
where, $t$ represents particle tunneling, $V$ is interaction between neighboring sites and $\mu$ is the chemical potential. The summation runs over all nearest neighbor sites $\langle jj'\rangle$ on a given geometry of the system. 
This Hamiltonian conserves the number of fermions as $[H,n]{=}0$, where $n{=}\sum_j n_j$. 
This implies that all the eigenstates of the Hamiltonian have a fixed number of fermions. In particular, the number of fermions $N_f$ in the ground state depends on $V$ which results in the filling factor $f{=}N_f/N$. 
As a further simplification, we assume that the sites are shallow, i.e. $\mu \ll t$, which is ensured by simply choosing $\mu {=} 0$ and $t {=} 1$. Note that while this parameter choice is not the most general, our focus here is to present a proof-of-principle improvement with the fermionic gates. Besides, for smaller atoms like Li, or relatively weak trapping laser fields, this assumption can be physically justified. It is also the regime where the MERGE protocol mentioned above is most successful. We now denote $H(t,V,\mu)$ in the equation above as simply $H(V)$. 

We simulate $N {=} 12$ sites arranged in three specific geometric configurations, a $1{\times}12$ Chain (Fig.~\ref{fig:spinless_chain}), a Ladder (Fig.~\ref{fig:spinless_ladder}) with $2$ rungs, i.e., $2{\times} 6$ system, and a $3{\times} 4$ Rectangular Lattice (Fig.~\ref{fig:spinless_rectangle}).  
Note that by spreading the system in two dimensions the entanglement in the ground state increases, which is known as area-law. Therefore, we expect that the simulation of such systems should be more resource-demanding than simple one-dimensional chains.

\subsection{Geometry 1: Open Chain}
\label{subsec:spinless_open_chain}
\begin{figure*}
    \centering
    \includegraphics[width=1\textwidth]{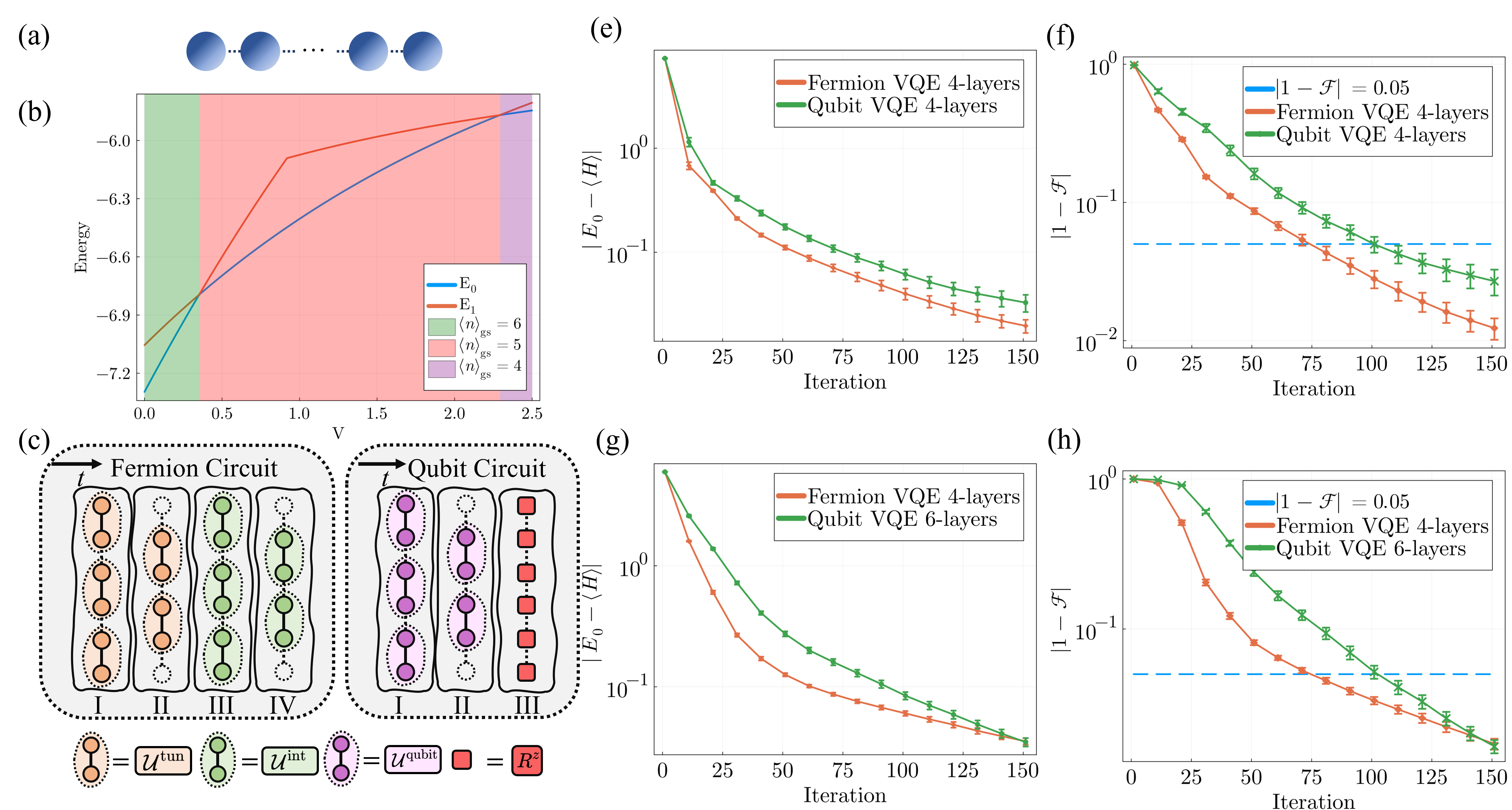}
    \caption{(a) Open chain configuration for $N{=}12$ spinless Fermi-Hubbard model. 
    (b) Dependence of particle number density on the Coulomb repulsion strength $V$ for the ground state. 
    (c) One layer of the fermionic circuit which is decomposed into four steps (orange steps denote $\mathcal{U}^{\text{tun}}_{jj'}$, green steps denote $\mathcal{U}^{\text{int}}_{jj'}$.). 
    (d) One layer of the qubit circuit which is decomposed into three steps (purple steps denote  $\mathcal{U}^{\text{qubit}}_{jj'}$, and red steps denote the phase rotations). (e) Performance of the fermionic circuit-based VQE (orange) vs the qubit circuit-based VQE (green) in terms of convergence to ground state energy for Fermi-Hubbard model interaction strength $ V= 0$. (f) Performance of the fermionic circuit-based VQE (orange) vs the qubit circuit-based VQE (green) in terms of fidelity with ground state for Fermi-Hubbard model interaction strength $ V= 0$. (g) Performance of the fermionic circuit-based VQE (orange) vs the qubit circuit-based VQE (green) in terms of convergence to ground state energy for Fermi-Hubbard model interaction strength $ V= 2$. (h) Performance of the fermionic circuit-based VQE (orange) vs the qubit circuit-based VQE (green) in terms of fidelity with ground state for Fermi-Hubbard model interaction strength $ V= 2$. Tunneling strength $t = 1$ throughout.}
    \label{fig:spinless_chain}
\end{figure*}

First, we consider the open boundary spinless fermion chain model (Fig.~\ref{fig:spinless_chain}(a)).  Physically, when $V{=}0$, the model reduces to the standard tight-binding model, and the ground state is at half-filling $f{=}1/2$. As the interaction $V$ increases, the corresponding energy penalty leads to a reduction of the fermion numbers $N_f$ in the ground state, one electron at a time, i.e., in discrete steps to $N_f{=}\frac{N}{2} -1, N_f{=}\frac{N}{2} {-}2$, and so on (see Fig.~\ref{fig:spinless_chain}(b)). For simulation, we consider two values of $V$ at $V{=}0$ and $V{=}2$, with the corresponding ground state fermion numbers being $N_f{=}6$ and $N_f{=}5$ respectively. 
The fermion and qubit circuit structure is shown in Fig.~\ref{fig:spinless_chain}(c) and (d), consisting of $\mathcal{U}^{\text{int}}_{jj'}$ and $\mathcal{U}^{\text{tun}}_{jj'}$ from Eq.~\eqref{eq:gen_bk}. 
In each circuit layer, the tunneling gates are grouped together which is then followed by interaction gates. 
There are two steps to fulfill either tunneling or interaction gates for all neighboring sites, resulting in a total depth of $4$ for each layer.

The simulation results, quantified by average energy $\langle H\rangle {=}\langle\psi(\vec{\theta})|H|\psi(\vec{\theta})\rangle$ and fidelity $\mathcal{F}{=}|\langle \psi(\vec{\theta})|E_0\rangle|^2$, for $V{=}0$ are depicted in Figs.~\ref{fig:spinless_chain}(e)-(f). 
Both fermion and qubit circuits contain four layers. As the figure shows, the fermion quantum simulator converges faster and shows smaller error bars. This means that fermionic simulators are not only faster but also more robust against random initialization. Note that non-interacting one-dimensional fermionic systems can be efficiently solved through Jordan-Wigner transformation on classical computers. Thus, aside from benchmarking, it is practically more important to investigate the non-solvable interacting fermions. We consider the interaction strength $V/t{=}2$ for which the filling factor of the ground state changes to $N_f{=}5$. 
The resulting average energy and fidelity are shown in  Figs.~\ref{fig:spinless_chain}(g)-(h), respectively. 
 Different from the non-interacting case, convergence is achieved with only $L{=}4$ layers for the fermionic circuit but requires $L{=}6$ layers for the qubit circuit. To achieve the threshold fidelity $\mathcal{F}{=}0.95$, the fermion VQE only needs $l_\text{I}{=}75$ iterations, while the qubit simulator requires $l_\text{I}{=}106$ iterations. 

Remarkably, in the interacting case, the improvement in convergence speed is even more significant than in the non-interacting case. The error bars still remain smaller for fermionic simulators. These results clearly show the superiority of fermionic quantum simulators over their qubit counterparts for simulating the fermionic many-body systems.

\subsection{Geometry 2: Ladder }
\label{subsec:spinless_ladder}
\begin{figure*}
     \centering
     \centering
     \includegraphics[width=\textwidth]{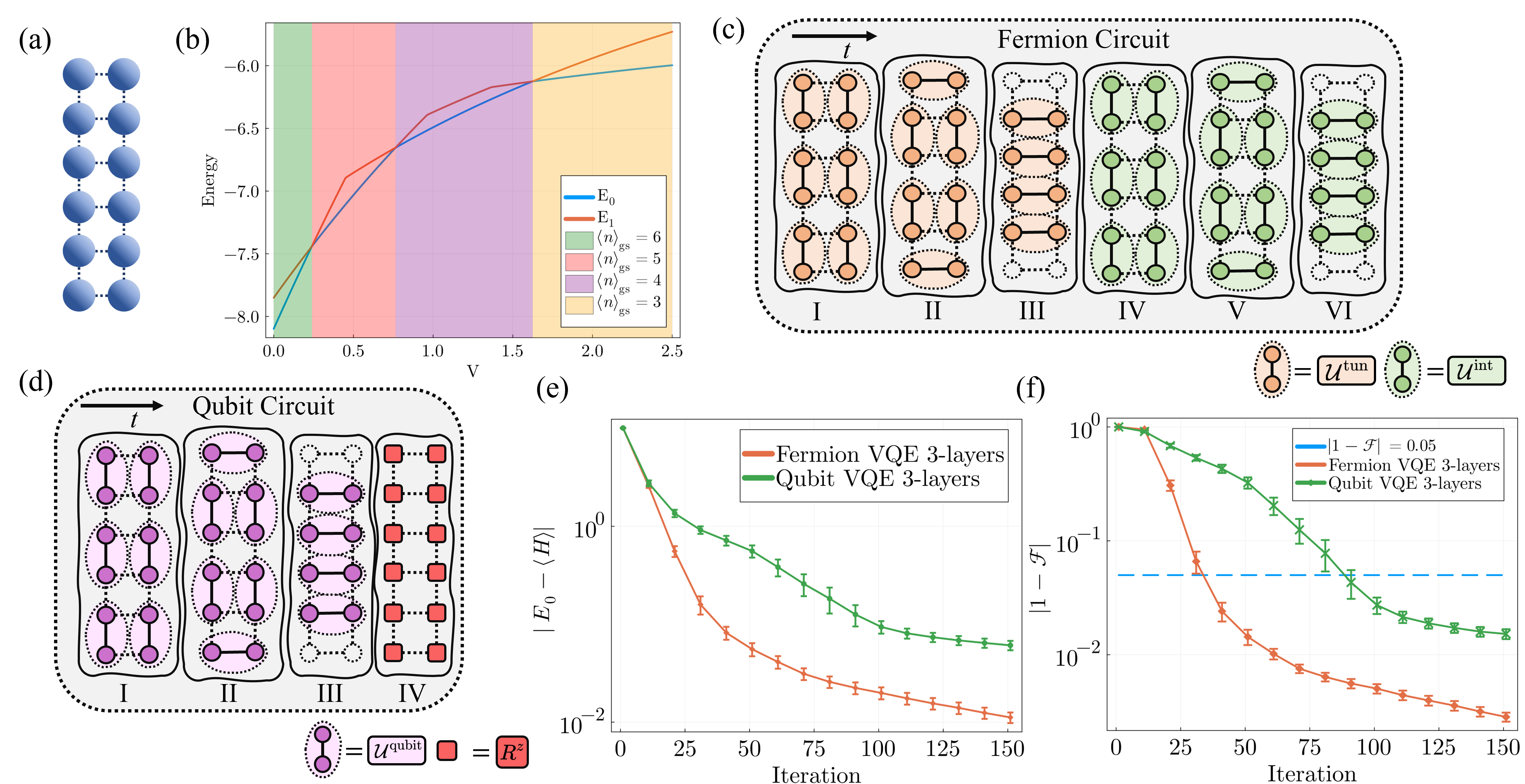}
    \caption{(a) Ladder $6{\times}2$ configuration for $N{=}12$ spinless Fermi-Hubbard model. (b) Dependence of particle number density on the Coulomb repulsion strength $V$ for the ground state. 
    (c) One layer of the fermionic circuit which is decomposed into $6$ steps (orange steps denote $\mathcal{U}^{\text{tun}}_{jj'}$, green steps denote $\mathcal{U}^{\text{int}}_{jj'}$.). 
    (d) One layer of the qubit circuit which is decomposed into $4$ steps (purple steps denote  $\mathcal{U}^{\text{qubit}}_{jj'}$, and red steps denote the phase rotations). (e) Performance of the fermionic circuit-based VQE (orange) vs the qubit circuit-based VQE (green) in terms of convergence to ground state energy for Fermi-Hubbard model interaction strength $ V = 2$. (f) Performance of the fermionic circuit-based VQE (orange) vs the qubit circuit-based VQE (green) in terms of fidelity with ground state for Fermi-Hubbard model interaction strength $ V = 2$. Tunneling strength $t = 1$ throughout.}
    \label{fig:spinless_ladder}
\end{figure*}

Now we consider the ladder model with two rungs, see Fig.~\ref{fig:spinless_ladder}(a),  as an intermediate between the 1D chain discussed above and the full 2D model.
As before, increasing the repulsive interaction $V$ leads to a decline in the ground state fermionic number density, as shown in Fig.~\ref{fig:spinless_ladder}(b). For designing the quantum circuit, we follow the same logic as 1D systems. In each circuit layer, we first group the tunneling gates $\mathcal{U}^\text{tun}_{jj'}$ together and then perform the interaction gate $\mathcal{U}^\text{int}_{jj'}$  in a similar fashion. Since there are more bonds in the ladder geometry the full operation of either tunneling or interaction gates can be fulfilled in three steps as shown in Fig.~\ref{fig:spinless_ladder}(c), resulting in total depth $6$ for each layer. 
The results, depicted in Figs.~\ref{fig:spinless_ladder}(e)-(f), show that the fermionic circuit converges far faster, requiring only $l_\text{I}{=}33$ iterations to reach a target fidelity of $\mathcal{F} {=} 0.95$, compared to $89$ iterations for the qubit circuit. As the results show by changing the geometry from a 1D chain to a ladder the improvement achieved by the fermionic simulator becomes even more pronounced. This is because the corresponding qubit Hamiltonian in the case of ladder is significantly more non-local which results in overhead resources. 
We note that both the qubit and fermionic circuits only need $L{=}3$ layers to converge within the allowed number of iterations,  which is less than $L{=}4$ layers required for the 1D chain previously. However, the circuit depth of each layer in this case is higher compared to the 1D chain and thus more gates are required. Again the error bars in fermionic simulators are smaller than the qubit simulators showing more robustness against the random initializations.

\subsection{Geometry 3: Rectangular Lattice}
\label{subsec:spinless_2d}
\begin{figure*}
     \centering
     \includegraphics[width=\textwidth]{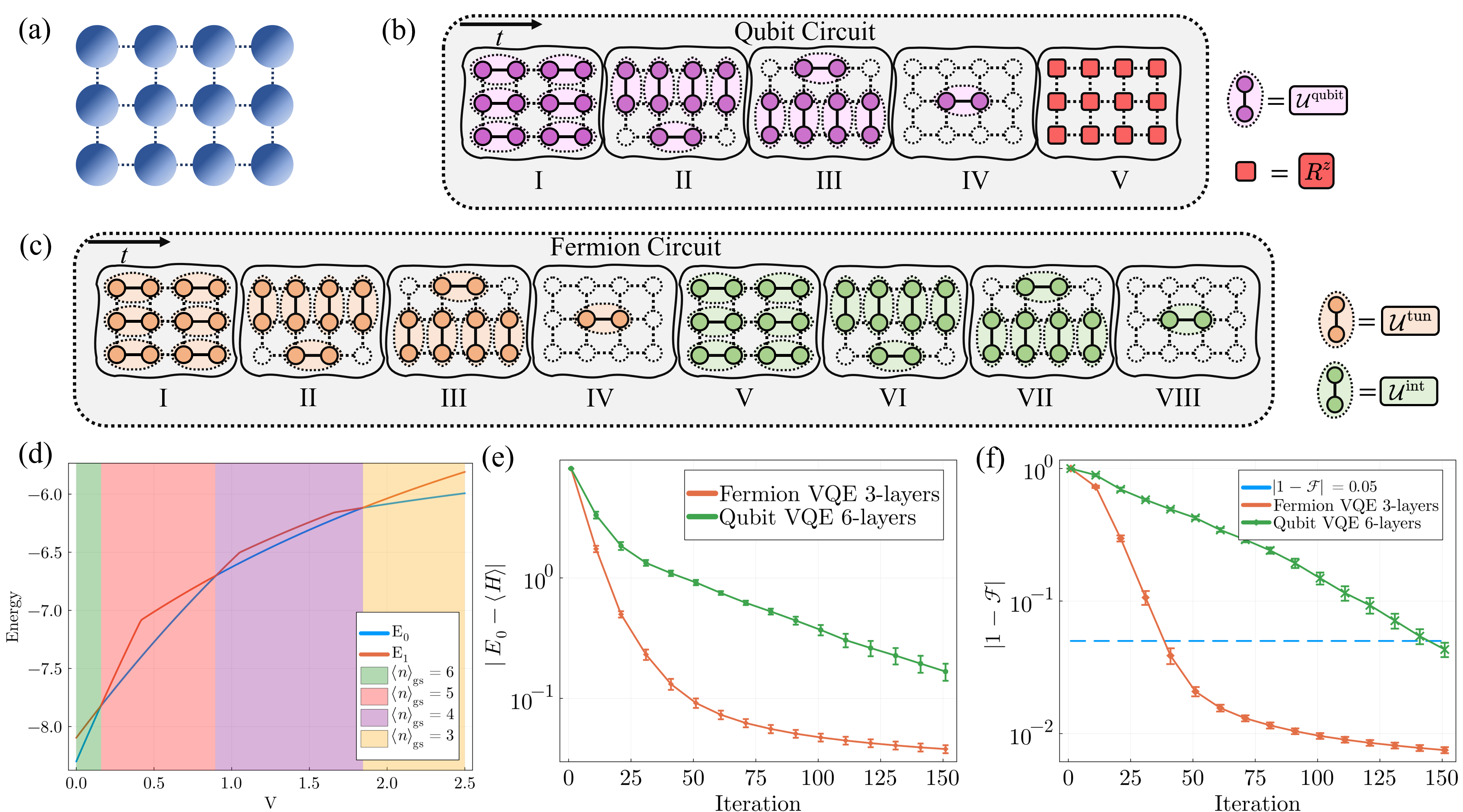}
   \caption{(a) Rectangular $3 \times 4$ configuration for $N{=}12$ spinless Fermi-Hubbard model. (b) One layer of the qubit circuit is decomposed into $5$ steps (purple steps denote  $\mathcal{U}^{\text{qubit}}_{jj'}$, and red steps denote the phase rotations).  (c) One layer of the fermionic circuit which is decomposed into $8$ steps (orange steps denote $\mathcal{U}^{\text{tun}}_{jj'}$, green steps denote $\mathcal{U}^{\text{int}}_{jj'}$).(d) Dependence of particle number density on the Coulomb repulsion strength $V$ for the ground state. 
   (e) Performance of the fermionic circuit-based VQE (orange) vs the qubit circuit-based VQE (green) in terms of convergence to ground state energy for Fermi-Hubbard model interaction strength $ V = 2$. (f) Performance of the fermionic circuit-based VQE (orange) vs the qubit circuit-based VQE (green) in terms of fidelity with the ground state for Fermi-Hubbard model interaction strength $ V = 2$. Tunneling strength $t = 1$ throughout.}
    \label{fig:spinless_rectangle}
\end{figure*}

As demonstrated above, by extension the geometry in 2D structures of the fermionic circuit shows a noticeable performance advantage over the qubit circuits.  We now consider a more pronounced  2D system, which is a $3{\times} 4$ rectangular lattice, shown in  Fig.~\ref{fig:spinless_rectangle}(a). In general, we expect the 2D system to be more challenging to simulate than the 1D system because of stronger entanglement and the presence of more couplings in the system.
As before, the filling factor of the ground state depends on the repulsive interaction $V$, as shown in Fig.~\ref{fig:spinless_rectangle}(d).  We again follow the same logic of grouping the gates in the circuit. For the rectangular lattice, we need four steps to perform either tunneling gates $\mathcal{U}^{\text{tun}}_{jj'}$ or interaction gates $\mathcal{U}^{\text{tun}}_{jj'}$ on all neighboring bonds, as shown in Fig.~\ref{fig:spinless_rectangle}(c). This means that each circuit layer has a depth of $8$. For converging to the ground energy, this fermionic circuit only needs three layers. 
In Figs.~\ref{fig:spinless_rectangle}(e)-(f), the average energies and the corresponding fidelities obtained by both fermionic and qubit simulators are shown. In order to converge to fidelity $\mathcal{F}{=}0.95$, one needs $L{=}3$ and $L{=}6$ layers in fermionic and qubit circuits, respectively. 
These results show that for higher dimensions, the gap in performance widens between fermionic and qubit circuit ansatzes.
As the figures show, for the qubit circuit, even with twice as many layers, it is still hard for it to converge to the true ground energy. In terms of fidelity, the qubit circuit needs $l_\text{I}{=}135$ iterations to achieve a fidelity of $\mathcal{F}{=}0.95$ but fermionic approach only needs $l_\text{I}{=}39$ iterations. 


Indeed, the VQE simulation of spinless fermionic Hubbard Hamiltonian shows that the fermionic circuit construction is superior to the qubit circuit construction for the simulation of spinless Hubbard Hamiltonians, with the performance advantage widening as we move towards higher dimensional systems, or indeed, for the same system with bigger interaction strengths. This can be seen more quantitatively in  the table \ref{table:spinless}, where we 
compare the performance of fermionic and qubit simulators in terms of both classical and quantum resources.

\begin{table*}[htbp]
\begin{tabular}{|c|cccccc|}
\hline
& \multicolumn{6}{c|}{ Spinless model $H(t{=}1,V{=}2)$}\\
\cline{2-7} 
\multirow{3}{*}{} & 
\multicolumn{2}{c|}{Chain $(1{\times} 12)$} & 
\multicolumn{2}{c|}{Ladder $(2{\times} 6)$} & 
\multicolumn{2}{c|}{Rectangle $(3{\times} 4)$} \\ 

\hline
Simulator & 
\multicolumn{1}{c|}{Fermion} & \multicolumn{1}{c|}{Qubit} & 
\multicolumn{1}{c|}{Fermion} & 
\multicolumn{1}{c|}{Qubit} & 
\multicolumn{1}{c|}{Fermion} & Qubit  \\ 

\hline
simulator size & 
\multicolumn{2}{c|}{$1{\times} 12$} & 
\multicolumn{2}{c|}{$2{\times} 6$} & 
\multicolumn{2}{c|}{$3{\times} 4$}\\ 
\hline
L & 
\multicolumn{1}{c|}{4} & 
\multicolumn{1}{c|}{6} & 
\multicolumn{1}{c|}{3} & 
\multicolumn{1}{c|}{3} & 
\multicolumn{1}{c|}{3} & 6      \\ 

\hline
$R_\text{Q}$ & 
\multicolumn{1}{c|}{88} & 
\multicolumn{1}{c|}{198} & 
\multicolumn{1}{c|}{96} & 
\multicolumn{1}{c|}{144} & 
\multicolumn{1}{c|}{102} & 306   \\ 
\hline

$l_\text{p}$          & 
\multicolumn{1}{c|}{176} & 
\multicolumn{1}{c|}{204} & 
\multicolumn{1}{c|}{192} & 
\multicolumn{1}{c|}{132} & 
\multicolumn{1}{c|}{204} & 276     \\ 

\hline
$l_\text{I}$ ($\mathcal{F}{=}0.95$) & 
\multicolumn{1}{c|}{75}      & 
\multicolumn{1}{c|}{106}   & 
\multicolumn{1}{c|}{33}      & 
\multicolumn{1}{c|}{89}    & 
\multicolumn{1}{c|}{39}      & 135    \\ 
\hline

$R_\text{C}$                     & 
\multicolumn{1}{c|}{13200}   & 
\multicolumn{1}{c|}{21624} & 
\multicolumn{1}{c|}{6336}    & 
\multicolumn{1}{c|}{11748} & 
\multicolumn{1}{c|}{7956}    & 37260 \\ 
\hline

\end{tabular}
\caption{Table for resource count of the simulation of various configurations of spinless fermionic Hubbard Hamiltonians with $N{=}12$ sites where the parameters are chosen to be $t{=}1$ and $V{=}2$.}
\label{table:spinless}
\end{table*}

\section{Simulating spinful Fermionic Many-Body systems}
\label{sec:IV}
In the previous section, we simulated the ground state of spinless fermionic Hubbard Hamiltonian and established the advantage gained through the use of fermionic quantum simulators. We now want to extend these results for the spinful Hubbard Hamiltonian, described as 
\begin{eqnarray}
    H = -t \sum_{\langle jj'\rangle}^{N}\sum_{\sigma = \uparrow, \downarrow} \left(c_{j \sigma}^{\dagger} c_{j'\sigma} + \text{h.c.} \right) -\mu \sum_{j} n_{j} \nonumber \\ + U \sum_{j} n_{j\uparrow} n_{j \downarrow} + V \sum_{\langle jj'\rangle} n_{j} n_{j'}\nonumber. \\
    \label{eq:hubbard}
\end{eqnarray}
As before, we again assume $t{=}1$ and $\mu{=}0$. Notice that since there are two spins, the intersite repulsion term $U$ is non-trivial between oppositely signed spins. We can then denote the full Hamiltonian $H(t,U,V,\mu)$ in terms of adjustable variables as $H(U,V)$. 

For the spinless system, each site only has two different states non-occupied or occupied. For $N$-sites, this information can be encoded in an $N$-qubit register. For the full description of spin $1/2$ fermions, each site has four possible configurations, non-occupied, occupied with one spin-up electron, occupied with one spin-down electron, and occupied with two opposite-spin electrons. For neutral atom arrays, this entails applying an external magnetic field to induce hyperfine level splittings. Therefore, for both qubit and fermionic simulators each site is encoded by two registers, one represents spin-up and one spin-down.

Before progressing further, let us describe how our circuit design is altered for the spinful fermionic case from the spinless case discussed in the last section. As mentioned before, there are two types of spinless gates in $\mathcal{G}$ set. For spinful fermions, the realizable gates depend on the details of trapping laser beams and pulses that one can apply. If one uses the same setup as Ref.~\cite{gonzalez2023fermionic} for realizing spinful fermions then the tunneling gate $\mathcal{U}_{jj'}^{\text{tun}}$ is replaced by two unitaries for each spins as
\begin{equation} \label{eq:U_tun_spinful}
\mathcal{U}_{jj',\alpha}^{\text{tun}}(\vec{\theta}) = e^{-i \left[ \frac{\theta_1}{2}\left( e^{-i\theta_2} c_{j,\alpha}^{\dagger} c_{j',\alpha}+ e^{i\theta_2} c_{j',\alpha}^{\dagger} c_{j,\alpha} \right) + \frac{\theta_3}{2} (n_{j,\alpha} - n_{j',\alpha}) \right]},
\end{equation}
where $\alpha{=}\uparrow{,} \downarrow$ represent the spin of the fermion. The interaction gate $\mathcal{U}_{jj'}^{\text{int}}$, however, is replaced by four different gates
\begin{equation} \label{eq:U_int_spinful}
\mathcal{U}_{jj',\alpha\alpha'}^{\text{int}}(\theta)=e^{-i\theta n_{j,\alpha}n_{j',\alpha'}},
\end{equation}
\noindent where, $\alpha,\alpha' \in \{\uparrow,\downarrow\}$ label the four possible interaction gates based on spin degrees of freedom. Based on the gate implementation of Ref.~\cite{gonzalez2023fermionic}, these gates can only be realized sequentially. Therefore, in our circuit design for spinful fermions, we also perform these gates sequentially. Note that by using alternative trapping methods and exploiting more Rydberg states one may be able to merge some of these unitaries and thus simplify the circuit. Nonetheless, for the sake of consistency, we stick to the operators given in Eqs.~(\ref{eq:U_tun_spinful}) and (\ref{eq:U_int_spinful}). 
\\
\noindent The benchmarking qubit circuit is built in analogy with the Fermionic circuits by retaining the same connections, and with weights defined by Eq.~\eqref{eq:xyz} earlier. 

\subsection{Geometry 1: Open Chain}
\label{subsec:spinful_open}
As mentioned above, in spinful systems with $N$ sites one has to use $2N$  registers.  
Hence, to keep the same simulator size as the spinless model, we have to cap the system size of the spin model at $N{=}6$ sites. Notice that when $N{=}6$, the simplest ladder example with two rungs coincides with the simplest 2D rectangular lattice of $2 {\times} 3 $ sites. We follow the exact same methodology as the spinless case.

\begin{figure*}
     \centering
     \includegraphics[width=\textwidth]{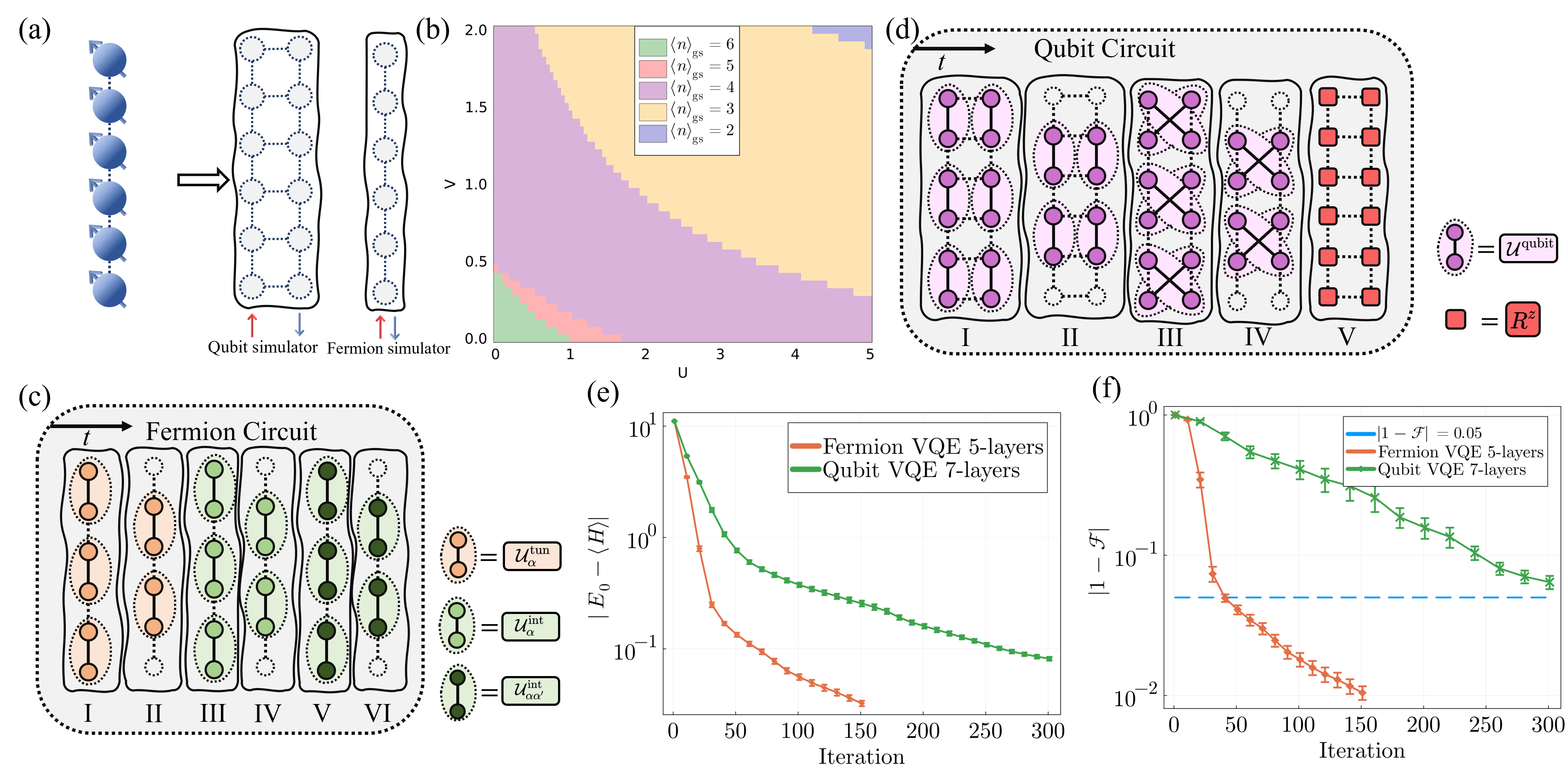}

   \caption{(a) Open chain configuration for $N{=}6$ spinful Fermi-Hubbard model as simulated by qubit (left) and fermonic (right) simulators.
    (b) Dependence of particle number density on the site depth $U$ and Coulomb repulsion strength $V$ for the ground state. 
    (c) One layer of the fermionic circuit which is decomposed into six steps (orange steps denote $\mathcal{U}^{\text{tun}}_{jj', \alpha \alpha}$, green steps denote $\mathcal{U}^{\text{int}}_{jj', \alpha \alpha'}$.). 
    (d) One layer of the qubit circuit which is decomposed into five steps (purple steps denote  $\mathcal{U}^{\text{qubit}}_{jj'}$, and red steps denote the phase rotations). (e) Performance of the fermionic circuit-based VQE (orange) vs the qubit circuit-based VQE (green) in terms of convergence to ground state energy for $ V= 0.5, U= 2.5$. (f) Performance of the fermionic circuit-based VQE (orange) vs the qubit circuit-based VQE (green) in terms of fidelity with ground state for $ V= 0.5, U= 2.5$. Tunneling strength $t = 1$ throughout.}
    \label{fig:spin_chain}
\end{figure*}

First, we consider the fermion spin chain model with size $N{=}6$ as shown in Fig.~\ref{fig:spin_chain}(a). The corresponding simulator structure is on the right. As we discussed, the simulator consists of $12$ registers, $6$ each for each spin configuration.  The number of fermions $N_f$ in the ground state for varying $V$ and $U$ is shown in Fig.~\ref{fig:spin_chain}(b). As $V$ or $U$ increases, the number of fermions declines from $6$ to $2$, step by step. In general, the onsite repulsion between two fermions is stronger than the coulomb repulsion between two neighbor sites, so we set $V{=}0.5$ and $U{=}2.5$. To simulate this spin model, the circuit of fermion VQE is designed as shown in Fig.~\ref{fig:spin_chain}(c), where tunneling gates only act on two neighbor spin-up sites or two neighbor spin-down sites in accordance with Pauli exclusion principle, as given by the hopping term of the Hamiltonian in Eq.~\eqref{eq:hubbard}.  in contrast, interaction gates are executed on the spin subspace at every site. Therefore, 
in each layer, for implementing tunneling gates two steps are needed while for interaction gates three steps. This makes the depth of a single layer equal to $L{=}5$. 
The numerical results for average energy and fidelity are shown in Figs.~\ref{fig:spin_chain}(e)-(f). The results are even more strongly in favor of the fermionic circuit than the spinless examples before. The qubit-based VQE converges far more slowly to the actual ground state even with $L{=}7$ layers, taking $l_\text{I}{=}325$ iterations to reach fidelity $\mathcal{F}{=}0.95$, while the fermionic simulators with $L{=}5$ layers converge to the ground state with the same fidelity $\mathcal{F}{=}0.95$ after only $l_\text{I}{=}73$ iterations. The quantum resource requirement, i.e., two-party elementary gate counts, is also more than halved by the fermionic circuit ($150$ vs. $420$ for the qubit circuit). 

\subsection{Geometry 2: Ladder}
\label{subsec:spin_ladder}
\begin{figure*}
     \centering
     \includegraphics[width=\textwidth]{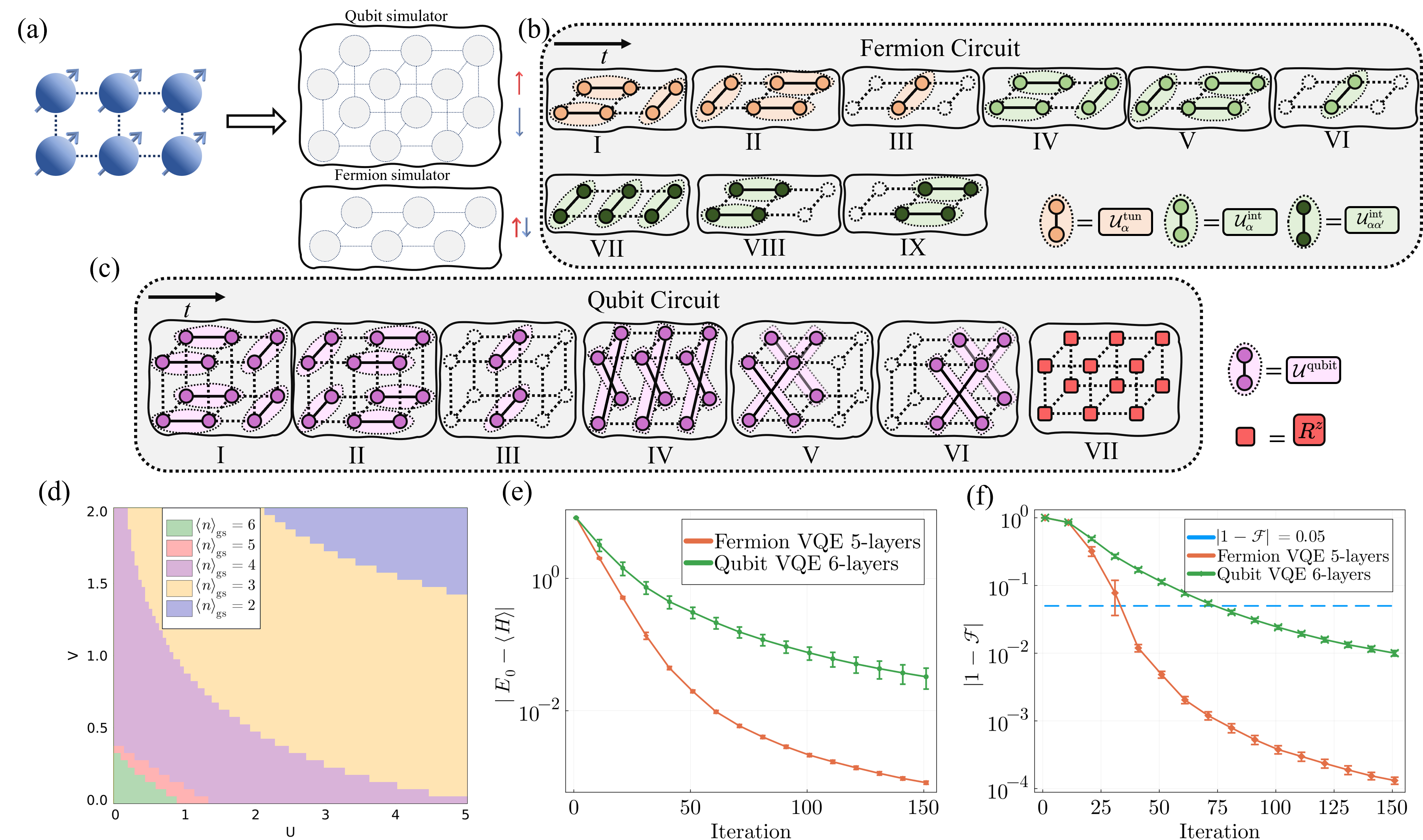}

   \caption{(a) Ladder $2{\times}3$ configuration for $N{=}6$ spinful Fermi-Hubbard model as simulated by qubit (top) and fermonic (bottom) simulators. (b) One layer of the fermionic circuit which is decomposed into $9$ steps (orange steps denote $\mathcal{U}^{\text{tun}}_{jj', \alpha \alpha}$, green steps denote $\mathcal{U}^{\text{int}}_{jj', \alpha \alpha'}$).(c) One layer of the qubit circuit is decomposed into $7$ steps (purple steps denote  $\mathcal{U}^{\text{qubit}}_{jj'}$, and red steps denote the phase rotations). (d) Dependence of particle number density on the site depth $U$ and Coulomb repulsion strength $V$ for the ground state. (e) Performance of the fermionic circuit-based VQE (orange) vs the qubit circuit-based VQE (green) in terms of convergence to ground state energy for $ V= 0.5, U= 2.5$. (f) Performance of the fermionic circuit-based VQE (orange) vs the qubit circuit-based VQE (green) in terms of fidelity with ground state for $ V= 0.5, U= 2.5$. Tunneling strength $t = 1$ throughout.}
    \label{fig:spin_ladder}
\end{figure*}

For a fermionic Hubbard model with $N{=}6$ sites, the only two-dimensional geometry is a $2{\times} 3$ ladder. The schematic of the system and its corresponding simulator with $12$ registers are shown in Fig.~\ref{fig:spin_ladder}(a). The number of fermions $N_f$ in the ground state of the system as a function of $U$ and $V$ is depicted in Fig.~\ref{fig:spin_ladder}(d).
To simulate this model, the corresponding one-layer quantum fermionic circuit is displayed in Fig.~\ref{fig:spin_ladder}(c). Similarly, the tunneling gates only act on sites that represent the same spins while interaction gates operate between all neighboring sites. Therefore, in a single circuit layer, the tunneling gates are realized through $3$ steps while interaction gates require $6$ steps to fulfill. This means that every circuit layer has a depth of $9$.  
The numerical results for the average energy and the corresponding fidelities on a qubit circuit with $L{=}6$ layers and fermionic circuits with $L{=}5$ layers are depicted in Figs.~\ref{fig:spin_ladder}(e)-(f). Again the results demonstrate that a fermionic circuit converges to fidelity $\mathcal{F}{=}0.95$ after $l_\text{I}{=}35$ iterations while the qubit circuit requires $l_\text{I}{=}74$ iterations.   This means that the fermionic simulators are more efficient with respect to both quantum and classical resources. 

\noindent In summary, the fermionic circuit is again, clearly better in the spinful case. The complete table for counting classical and quantum resources for each simulation is provided in Table~\ref{table:spinful}, which quantitatively backs up this assertion.

\begin{table*}[t]
\begin{tabular}{|c|cccc|}
\hline
&\multicolumn{4}{c|}{Spin model $H(t{=}1,U{=}2.5,V{=}0.5)$ }\\ 
\cline{2-5} 
\multirow{3}{*}{} &  
\multicolumn{2}{c|}{Chain $(1{\times} 6)$}  & 
\multicolumn{2}{c|}{Ladder/Rectangle $(2{\times} 3)$} \\ 
\hline
Simulator     & 
\multicolumn{1}{c|}{Fermion} & 
\multicolumn{1}{c|}{Qubit} & 
\multicolumn{1}{c|}{Fermion} & 
\multicolumn{1}{c|}{Qubit} \\ 
\hline
simulator size &  
\multicolumn{1}{c|}{$6{\times} 1$} & 
\multicolumn{1}{c|}{$6{\times} 2$} & 
\multicolumn{1}{c|}{$2{\times} 3$} &
\multicolumn{1}{c|}{$2{\times} 3{\times} 2$}  \\ 
\hline
$L$        &        
\multicolumn{1}{c|}{5} & 
\multicolumn{1}{c|}{7} & 
\multicolumn{1}{c|}{5} & 6     \\ 
\hline
$R_\text{Q}$  & 
\multicolumn{1}{c|}{150} & 
\multicolumn{1}{c|}{420} & 
\multicolumn{1}{c|}{210} & 504   \\ 
\hline
$l_\text{p}$           & 
\multicolumn{1}{c|}{250} & 
\multicolumn{1}{c|}{364} & 
\multicolumn{1}{c|}{350} & 408   \\ 
\hline
$l_{\text{I}}$ ($\mathcal{F}{=}0.95$)                 
& \multicolumn{1}{c|}{73}      
& \multicolumn{1}{c|}{325}     
& \multicolumn{1}{c|}{35}     & 74    \\ 
\hline
$R_\text{C}$   &  
\multicolumn{1}{c|}{18250}   & 
\multicolumn{1}{c|}{118300}   & 
\multicolumn{1}{c|}{12250}   & 30192 \\ \hline
\end{tabular}
\caption{Table for resource count of the simulation of various configurations of spinful fermionic Hubbard Hamiltonians with $N{=}6$ sites where the parameters are chosen to be $t{=}1$, $U{=}2.5$ and $V{=}0.5$.}
\label{table:spinful}
\end{table*}

\subsection{Resource Efficiency}
\label{subsec:spinful_efficiency}
Table~\ref{table:spinful} details both classical and quantum resources utilized by qubit and Fermionic simulators for various configurations of the spinful Hubbard Hamiltonian. Even more prominently than the spinless case, the Fermionic simulator consumes significantly less quantum resource $R_\text{Q}$ and classical resource $R_\text{C}$. As an illustrative example, We could only converge the qubit simulator for the open chain to $95\%$ fidelity after  $325$ iterations, however, the Fermionic simulator converges to this target fidelity nearly five times faster ($73$ iterations) and with shallower circuits (only $5$ layers vs $7$ layers for the qubit), thus proving its superiority.

\begin{figure}[h!]
     \centering
     \includegraphics[width=0.5\textwidth]{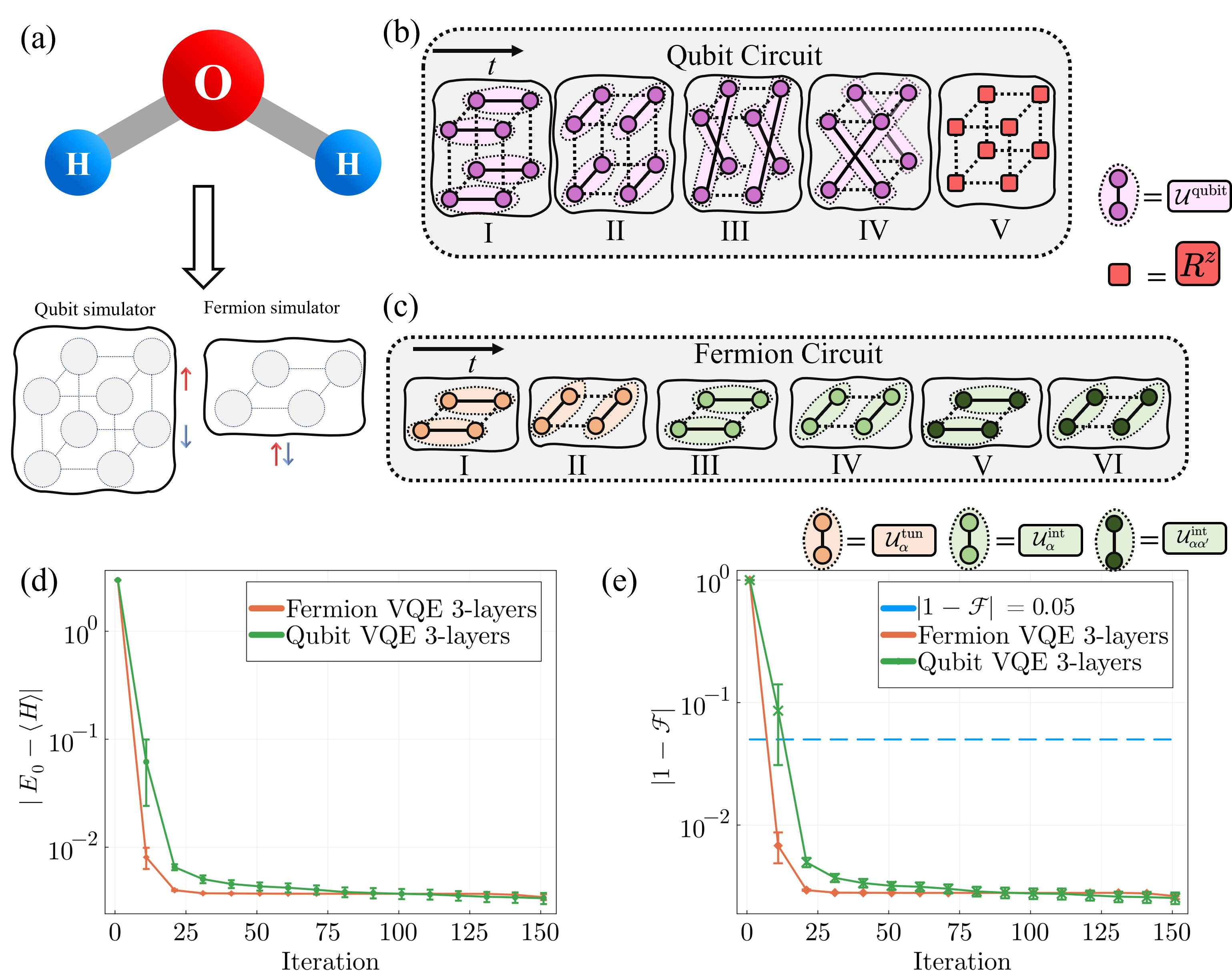}
    \caption{(a) The molecular geometry of water with $4$ spin orbits,
    (b) One layer of the qubit circuit decomposed into $5$ steps (purple steps denote $\mathcal{U}^{\text{qubit}}_{jj'}$, red denote phase rotation $R_j^z$.)
    (c) One layer of the fermionic circuit decomposed into $7$ steps (orange steps denote $\mathcal{U}^{\text{tun}}_{jj'}$, green steps denote $\mathcal{U}^{\text{int}}_{jj'}$). 
    (d) Performance of the fermionic circuit based VQE (orange) vs. the qubit circuit-based VQE (green), in terms of convergence to ground state energy. (e) Performance of the fermionic circuit-based VQE (orange) vs. the qubit circuit-based VQE (green), in terms of fidelity to the actual ground state. }
    \label{fig:water}
\end{figure}

\section{Resource Efficiency and Scalability}
\label{sec:scalability}

Tables ~\ref{table:spinless} and ~\ref{table:spinful} detail both classical and quantum resources utilized by qubit and Fermionic simulators for various configurations of the spinless and spinful Hubbard Hamiltonians. The quantum resource counts $R_\text{Q}$, i.e., the two-body gate count for the whole circuit, increases as the systems acquire more and more width, for both Fermionic and qubit architectures. However, unlike the Fermionic case, the qubit circuit requires a lot more depth to converge for the 2D lattice. Also, while the quantum resource required increases for the 2D lattice, the open chain actually requires the most classical resource for Fermionic simulators, which is in contrast with the qubit simulator.

\begin{figure}
     \centering
     \includegraphics[width=0.5\textwidth]{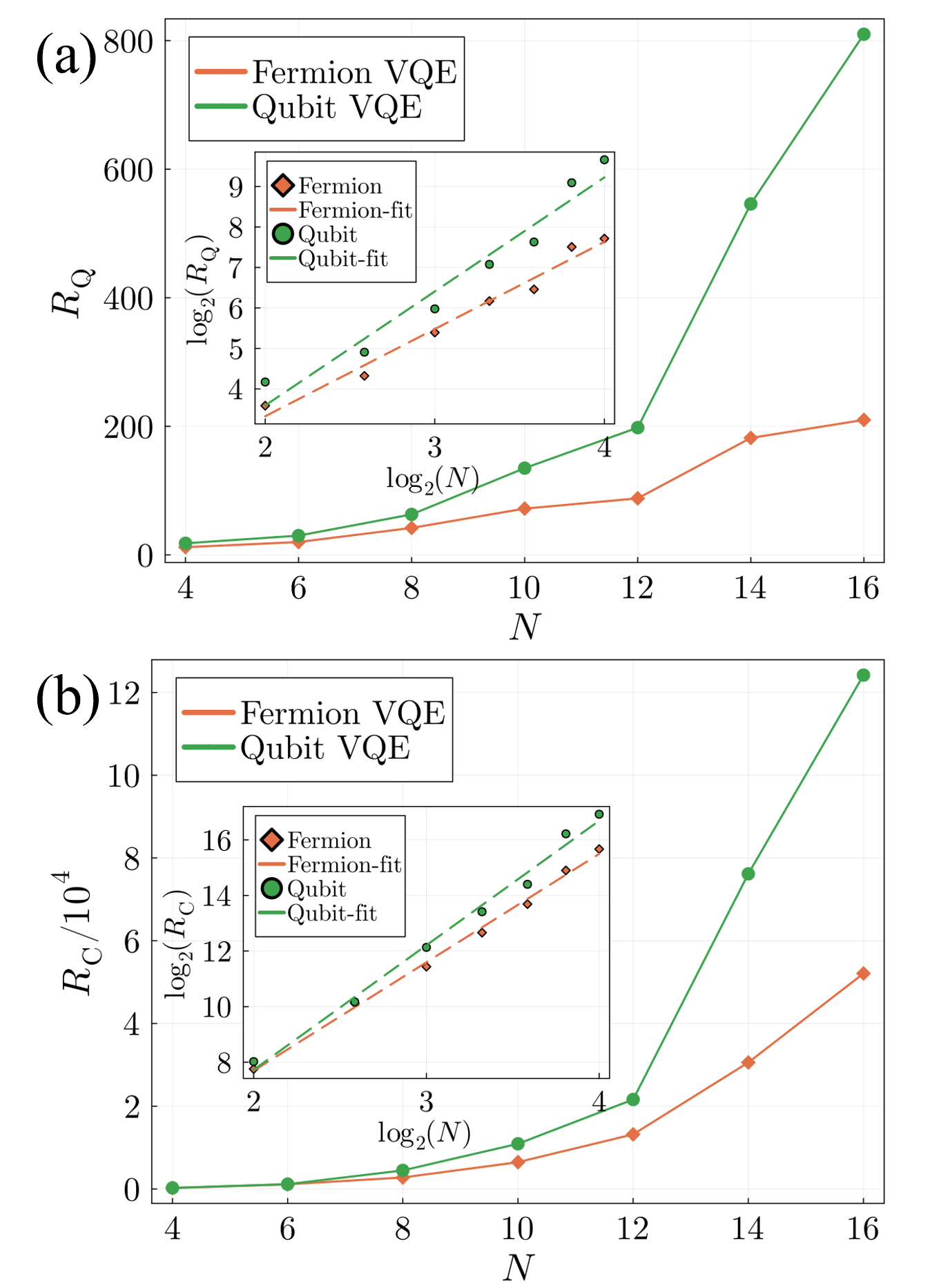}
        \caption{(a) Required quantum resource $R_\text{Q}$, and (b) the corresponding classical resource $R_\text{C}$, as system size $N$ increases for the spinless 1D open chain. Orange lines denote fermionic based and green lines denote qubit-based simulator performance. Inset to Fig.~(a) depicts fitting lines for $\log R_\text{Q}$ vs $\log N$ (slopes $\approx 2.19$ for Fermionic vs. $\approx 2.81$ for qubit ). Inset to Fig.~(b) depicts fitting lines for $\log R_\text{C}$ vs $\log N$ (slopes $\approx 3.90$ for Fermionic vs. $\approx 4.48$ for qubit).  Interaction strength $V = 2$ in every case. Tunneling strength $t = 1$ throughout.}
        \label{fig:spinless_scale}
\end{figure}

As demonstrated above, the fermionic simulator showed significant advantages over the qubit-based simulator on both quantum and classical resources. From table~\ref{table:spinless} and table~\ref{table:spinful} for spinless and spinful cases respectively, we know these advantages grow bigger with the spatial dimension of the system. The important question is - \emph{do these advantages scale up when the system size $N$ increases ?}. This is crucial for a NISQ quantum simulator because of limited resources available. Here, we take the counts of $R_\text{Q}$ and $R_\text{C}$ to compare the resource consumption of qubit- and fermionic-based quantum simulators.

\textcolor{black}{We note from our data that the resource requirements grow polynomially with system size $N$, i.e., $R_\text{Q} \sim N^{\beta_Q}$ and $R_\text{C} \sim N^{\beta_C}$, with scaling exponents $\beta_Q$ and $\beta_C$ respectively. As an illustration, demonstrated in Fig~\ref{fig:spinless_scale}, our data for the spinless 1D open chain indicates that $\beta_Q = 2.19, \beta_C = 3.90$ for the fermionic circuit vs $\beta_Q = 2.81, \beta_C = 4.48$ for the qubit circuit. While the exponents quoted above are not conclusive since we had to confine ourselves to short chains, they nonetheless confirm that vis-a-vis the qubit simulator, the fermionic simulator shows more resource efficiency for both quantum and classical parts of the protocol as $N$ increases, i.e., the relative advantage with respect to the qubit simulator is scalable. In Table~\ref{table:scalability}, we list the scaling exponents for other geometries studied in the paper. Due to computational limitations, finding meaningful scaling exponents becomes difficult for thicker ladders with more rungs or especially for qubit architectures,  particularly in all except the simplest 1D chain for the spinful case, one can nonetheless draw two tentative conclusions from Table~\ref{table:scalability}. Firstly, for all the configurations studied, both classical and quantum resource scaling significantly favors the fermionic architecture in every case. Secondly, the qubit architecture in the spinful case is significantly less scalable than the corresponding 1D spinless chain with the corresponding scaling exponents being larger for the spinful $(\beta_C =5.86, \beta_{Q} = 3.24)$ than the spinless $(\beta_C = 4.48, \beta_Q = 2.81)$. However, for the fermionic architecture, the simulation of the spinful chain is roughly as scalable as the spinless in terms of gate counts ($\beta_Q = 2.23$ for the spinful chain vs $\beta_Q = 2.19$ for the spinless chain) and the spinful chain actually shows better simulation scalability in terms of classical optimization resource requirements ($\beta_C = 3.35$ for the spinful chain vs $\beta_C = 3.90$ for the spinless chain). Together, they indicate that given the severe constraints of the NISQ era, Fermionic simulators can allow us to tackle systems, especially spinful systems, far larger than the existing qubit simulators can.}

\begin{table}[]
\begin{tabular}{|c|cccc|}
\hline
                                   & \multicolumn{4}{c|}{{\color[HTML]{000000} Spinless model}}                                                                                        \\ \hline
                                   & \multicolumn{2}{c|}{$\beta_C$}                                              & \multicolumn{2}{c|}{$\beta_Q$}                                              \\ \hline
Configurations                           & \multicolumn{1}{c|}{Fermion}       & \multicolumn{1}{c|}{Qubit}         & \multicolumn{1}{c|}{Fermion}       & Qubit                              \\ \hline
$N \times 1$                       & \multicolumn{1}{c|}{3.90$\pm$0.11} & \multicolumn{1}{c|}{4.48$\pm$0.19} & \multicolumn{1}{c|}{2.19$\pm$0.12} & 2.81$\pm$0.28                      \\ \hline
$N \times 2$                       & \multicolumn{1}{c|}{3.36$\pm$0.11} & \multicolumn{1}{c|}{4.12$\pm$0.10} & \multicolumn{1}{c|}{2.12$\pm$0.10} & 2.23$\pm$0.12                      \\ \hline
$N \times 3$                       & \multicolumn{1}{c|}{3.76$\pm$0.28} & \multicolumn{1}{c|}{4.88$\pm$0.28} & \multicolumn{1}{c|}{2.49$\pm$0.09} & 2.75$\pm$0.11                      \\ \hline
$N \times 4$                       & \multicolumn{1}{c|}{3.69$\pm$0.17} & \multicolumn{1}{c|}{4.67$\pm$1.00} & \multicolumn{1}{c|}{2.06$\pm$0.13} & 2.40$\pm$0.06                      \\ \hline
$N \times N$                       & \multicolumn{1}{c|}{3.77$\pm$0.14} & \multicolumn{1}{c|}{5.84}          & \multicolumn{1}{c|}{2.31$\pm$0.28} & 3.06                               \\ \hline
\multicolumn{1}{|l|}{}             & \multicolumn{4}{c|}{Spinful model}                                                                                                                \\ \hline
$N \times 1$ & \multicolumn{1}{l|}{3.35$\pm$0.25} & \multicolumn{1}{l|}{5.86$\pm$0.50} & \multicolumn{1}{l|}{2.23$\pm$0.22} & \multicolumn{1}{l|}{3.24$\pm$0.19} \\ \hline
\end{tabular}
\caption{Table for scaling exponents of $N$ for various configurations of the Hubbard Hamiltonian for Fermionic and Qubit simulators}
\label{table:scalability}
\end{table}

\section{Quantum Chemistry : Simulation of \ch{H2O} Molecule}
\label{sec:V}
In the previous sections, we discussed the use of direct fermionic circuits to implement the variational algorithm for obtaining the ground state of spinless or spinful fermionic many-body models with several different geometric configurations. These results all point towards a significant advantage being obtainable in terms of both classical and quantum resource efficiency than the standard Qubit-based VQE schemes.

In this section, we consider an example from the domain of quantum chemistry to illustrate the advantage of Fermionic simulators. We remember that chemical bonds happen because interacting electrons, each delocalized between different atomic orbitals, are in a bound state, i.e., have a negative ground state energy. This molecular Hamiltonian can thus be written as a many-electron Fermionic Hamiltonian where in the second-quantized notation, $c_i (c_i^{\dagger})$ are lowering and raising Fermionic operators for the $i$-th orbital. The ground state energy of the molecular Hamiltonian then gives an idea of the strength of the bond. Thus, qubit-based quantum simulators again encounter the same problem of fermion-to-qubit encoding-induced overheads, and we expect the Fermionic simulator to be significantly better. In fact, in Ref.~\cite{gonzalez2023fermionic}, two four-body generalized gates corresponding to the terms in coupled cluster ansatz of approximating the molecular Hamiltonian \cite{anand2022a} were constructed from the repeated application of Fermionic gates  $\mathcal{U}^{\text{tun}}_{jj'}$ and  $\mathcal{U}^{\text{int}}_{jj'}$. With the help of these gates, Ref.~\cite{gonzalez2023fermionic} found the ground state of \ch{LiH} as a case study. 
Here, we provide an explicit demonstration of the efficiency of the fermionic simulator by finding the ground energy of a molecule of water, whose molecular structure is shown in Fig.~\ref{fig:water}(a) \footnote{The molecular Hamiltonian of $\ch{H2O}$ is created by OpenFermion package.}. For the $\ch{H2O}$ molecule, we have $4$ active orbitals and $4$ electrons hopping between them. The simulator has to, therefore, contain $8$ registers to account for spin degrees of freedom. Different from the approach of \cite{gonzalez2023fermionic}, we do not explicitly use the four-body Fermionic gates as building blocks of the unitary, instead using two-body tunneling and interaction gates introduced earlier. One layer of the fermionic quantum circuit is shown in Fig.~\ref{fig:water}(c), which shows that the tunneling gates can be implemented in $2$ steps while the interaction gates require $3$ steps, therefore, a single layer of the circuit has the depth of $5$. The results of average energy and fidelity are shown in Figs.\ref{fig:water}(d)-(e), respectively. Again, compared with qubit circuits, fermionic circuits outperform qubit-based circuits in terms of convergence speed ($l_\text{I}= 8$ for Fermionic vs. $13$ for qubit, less is better), and quantum resources ($R_\text{Q}= 72$ for Fermionic vs. $144$ for qubit, less is better). In terms of classical resource cost, both are roughly similar ($R_\text{C}= 1280$ for Fermionic vs. $1560$ for qubit, less is better), and the error bars representing variability of output on randomized initializations are narrower for the Fermionic circuit than the qubit circuit. Thus, even for the generic Fermionic circuit, an advantage over the qubit paradigm is present. Please note that Fig.~\ref{fig:water}(d)-(e) are in the semilog scale and they also indicate the onset of the barren plateau problem at larger iterations, albeit beyond our fidelity accuracy threshold.
 
Let us also note that we have only considered the total particle number conservation property in this algorithm. Incorporating all available symmetries of molecules, such as the total spin, time-reversal, and point-group and crystallographic symmetries, into the construction of fermionic quantum simulators should further reduce the required resource costs and improve the performance of the fermion simulation algorithms even more. 
\\
\section{Conclusion}
\label{sec:conclusion}
Efficient simulation of strongly interacting fermionic systems is crucial in every physical and technological domain. Shortcomings of classical computation are long known for the simulation of quantum systems, which leaves quantum simulators as the only viable solution. Quantum simulators are rapidly emerging in various physical platforms, with qubit architectures becoming the default arrangement. However, the imperfect nature of NISQ simulators prohibits universal large-scale quantum computation, which makes resource efficiency a critical issue for the foreseeable future. Simulation of fermionic systems on qubit-based quantum simulators requires fermion-to-qubit mappings which introduce additional resource overhead, limiting the scalability to larger systems and introducing new challenges for trainability. 
Analog fermionic quantum simulators~\cite{spar2022realization,yan2022two} have been experimentally demonstrated in neutral atom arrays, paving the way for the realization of programmable fermionic digital simulators. Moreover, in a recent parallel work \cite{wei2023hubbard}, the inverse problem of setting the optical tweezer trap configuration for various Hubbard model configurations was also treated. In this paper, we compare the performance of fermionic- and qubit-based simulators for VQE simulations of the fermionic many-body ground state in both condensed matter systems and quantum chemistry problems. Our results show that fermionic quantum simulators offer a clear, scalable, and significant advantage in terms of both classical as well as quantum resources. In addition, fermionic simulators show less sensitivity to random initializations of circuit parameters. Thus, our work opens up the possibility of simulating much bigger fermionic systems in future, which can potentially have a big impact on material design and quantum chemistry.\\

\section*{Acknowledgements}
Authors acknowledge support from the National Key R\&D Program of China (Grant No. 2018YFA0306703), the National Science Foundation of
China (Grants No. 12050410253, No. 92065115, and No. 12274059), and the Ministry of Science and Technology of China (Grant No. QNJ2021167001L). 


\bibliography{apssamp}

\end{document}